\documentstyle[aps, prb]{revtex}

\input{tcilatex}

\begin{document}
\title{DIVERGENCE OF THE QUASIPARTICLE LIFETIME WITH DOPING AND EVIDENCE FOR
PRE-FORMED PAIRS BELOW T* IN YBa$_{2}$Cu$_{3}$O$_{7-\delta }$ : Direct
measurements by femtosecond time-resolved spectroscopy.}
\author{D.Mihailovic, B.Podobnik, J.Demsar,}
\address{Jozef Stefan Institute, 10001 Ljubljana, Slovenia}
\author{G.Wagner and J.Evetts}
\address{Materials Department, University of Cambridge, Cambridge, U.K.}
\maketitle

\begin{abstract}
We report new time-resolved data of quasiparticle relaxation and Cooper pair
recombination dynamics in YBa$_{2}$Cu$_{3}$O$_{7-\delta }$, measured as a
function of temperature and doping $\delta $ using femtosecond optical
spectroscopy. The data show the existence of a normal state pseudo-gap for
in-plane charge excitations below $T^{*}$ and an unusual divergence of the
quasiparticle relaxation time in the superconducting state $\tau _{s}$ with $%
\delta $ towards optimum doping. In the underdoped state, no change in the
amplitude of the induced transmission (which is proportional to the DOS at $%
E_{F}$), or relaxation time $\tau _{s}$ is observed at $T_{c}$. From the sum
rule, both observations signify that no gap opens at $T_{c}$ in underdoped
YBCO for $\delta >0.15.$ $T_{c}$ in this case only signifies the onset of
phase coherence. The presented data thus suggest pair formation with an
associated redistribution of the DOS starting at $T^{*}$ and the
establishment of phase coherence at $T_{c}$ consistent with Bose-Einstein
condensation. In the optimally doped material, $\delta $ $\approx $ 0.1, on
the other hand, both a divergence in lifetime and a change of the DOS\ occur
at $T_{c}$, signifying the opening of a gap and the occurence of pairing
takes place simultaneously.
\end{abstract}

\newpage

\section{Introduction}

The low-energy electronic excitation spectrum in high-temperature
superconducting cuprates has been controversial ever since the first
experimental data were reported on the subject more than 10 years ago. The
standard measurement techniques used to obtain experimental data have been
mainly infrared reflectivity, transmission and ellipsometry together with
electronic Raman scattering. Although there was significant agreement
regarding the raw data between different groups\cite{Ginsberg}, controversy
arose when interpretation of the spectra was attempted. The single particle
spectrum on the other hand was very successfully investigated by
angle-resolved photoemission (ARPES)\cite{ARPES dispersion}. However in this
case, just as for optical spectroscopy, the complexity of the observed
multi-component spectral features necessitates the use of a model in
interpreting the data. Furthermore in all time-integrated spectroscopies the
issue of whether the spectrum shows homogeneous or inhomogeneous linewidth
cannot be easily circumvented, so particle lifetimes determined by these
techniques are necessarily ambiguous.

In this paper we describe an example of the application of time-resolved
optical spectroscopy to the direct investigation of low-energy electronic
excitation dynamics. The method by default directly gives quasiparticle
lifetimes in the case of high-temperature superconductor materials. From
such measurements as a function of temperature and doping $\delta $ in YBa$%
_{2}$Cu$_{3}$O$_{1-\delta }$ we are able to infer the occurrence of
Bose-Einstein condensation of quasiparticles at $T_{c}$ in the underdoped
state and a cross-over to BCS-like state near optimum doping.

\section{Experiments}

The success of the time-resolved spectroscopy method relies on two factors.
Firstly the technological developments of femtosecond lasers and related
technology, with the recent development of high-frequency lock-in
amplifiers. Secondly - a particularly important generic feature of the
cuprate materials - the existence of a charge-transfer resonance in the
wavelength region of 800 nm easily accessible by these lasers. The
assignment of this resonance by use of laser wavelengths ranging from 1500
to 350 nm has been particularly important for the understanding the
time-resolved measurements\cite{Stevens}. In previous work\cite{Han}$^{,}$%
\cite{tr refs}, the spectroscopy off-resonance (usually at 2 eV) showed
features which were substantially more difficult to interpret.

The laser used in the present time-resolved experiment was a Ti:Sapphire
mode-locked laser giving 150 fs pulses at 800 nm with a repetition rate of
88 MHz. The pump pulse train average power was typically between 10 and 120
mW, while the probe pulse train was typically 0.2 mW or less. The two beams
were focussed onto the sample mounted onto a copper block in an Oxford
Instruments Microstat. The pump beam was modulated at 200 kHz and the probe
detected via an amplified photodiode and a high-frequency digital lock-in
amplifier (EG\&G 7260).

For the understanding of the present experiments, it is important to
establish that the optical transitions involve excitations from the ground
state (of predominantly O character) near $E_{F}$ to unoccupied states (of
predominantly Cu orbitals) 1.5 eV above in energy. In such a resonance case,
the changes in the ground state DOS occurring with temperature, particularly
at $T_{c}$ and $T^{*}$ can thus be measured by pump-probe spectroscopy. The
evidence for the assignment given above is summarized below. From early
X-ray work of Bianconi and others\cite{Bianconi88} it emerged that the
charge carriers (holes) are located mainly on the O ions, while the Cu $d$
(upper Hubbard) band is approximately 1.5-1.8 eV\ higher in energy
(depending on the material and level of doping). Experiments on YBa$_{2}$Cu$%
_{3}$O$_{x}$ for example have shown that the charge transfer (CT) transition
between O and Cu is strongly observed in optical conductivity, optical
ellipsometry\cite{Kircher LDA} and absorption spectra\cite{PC} as well as
photoconductivity\cite{PC} at 1.8 eV with a typical width of 0.1 eV. Upon
doping it splits into two bands approximately $\sim 0.1$ eV apart at the
insulator-to metal transition\cite{Kircher LDA}$^{,}$\cite{Demsar} and
eventually at optimum doping only a peak at 1.5 eV is clearly visible. A
feature at 1.5 eV is clearly observed also in large-shift electronic Raman
scattering\cite{Cooper}with an apparent splitting observed at $\delta $ $%
\sim $ 0.6. This probably indicates that the local symmetry for the CT\
transition is lower than orthorhombic and has no center of inversion,
otherwise the 1.5 eV feature would not be seen both in {\it gerade} (Raman)
and {\it ungerade} (infrared)\ transitions.

Further confirmation that the initial states for optical excitations at 1.5
eV cross $E_{F}$ comes from thermal differential reflectance (TDR)\cite
{Little} and time-resolved (TR) optical spectroscopies\cite{Stevens}. The
former measures the change in reflectivity which occurs at $T_{c}$ as a
result of the redistribution of the DOS\ and shows distinctly a double peak
at 1.5 eV similar to the one seen in the optical conductivity and Raman
spectra. TR spectroscopy shows the same feature at 1.5 eV in the optimally
doped YBa$_{2}$Cu$_{3}$O$_{7-\delta }$ - albeit with much lower resolution -
also appearing below $T_{c}$\cite{Stevens} and is similarly thought to be
caused by a redistribution of initial states for the optical transitions at $%
E_{F}$. In this case the 4 : 1 in-plane to out-of-plane polarization ratio
also confirms this assignment. Yet more independent confirmation that the
initial states for absorption are at $E_{F}$ come from ARPES\cite{ARPES
dispersion}\ in conjunction with theoretical calculations, while the effect
of doped holes on the 1.5 eV transition was nicely demonstrated by Matsuda
et al\cite{Matsuda}.

The schematic diagram for the resonant pump and probe processes which is
based on the assignment discussed above is shown in Figure 1. The carriers
are first excited by the pump pulse from occupied states at or below $E_{F}$%
. The second, probing step involves determining - as a function of time -
the resulting changes in the density of states at $E_{F}$ $N(0),$ caused by
the photoexcited quasiparticles by measuring the change in transmission $%
{\cal T}$ of the sample with a delayed weak laser probe pulse. In the
adiabatic approximation, the golden rule gives the {\em change }in optical
absorbance of the probe pulse to be proportional to the change in the DOS at 
$E_{F}$ i.e. $\delta A$ $\propto \delta N(0)$ caused by the pump pulse.
Since - as was discussed above - this is in turn proportional to $N(0)$, the
time-resolved changes in optical transmission $\delta {\cal T}/{\cal T}%
\,(=-\delta {\cal A}/{\cal A})$ can directly probe temperature- and $\delta $%
- dependence of $N(0)$. The probe pulse was polarized in the $a-b$ plane to
probe predominantly CuO-plane excitations (the signal was found previously
to be independent of pump polarization\cite{Stevens}).

\section{Experimental results}

The induced transmission $\Delta {\cal T}/{\cal T}$ following the pump pulse
typically shows a relatively fast relaxation of $\tau \sim $ 0.5 - 3 ps,
followed by a distinct long-lived component with a decay time 10 ns or more%
\cite{Stevens,Thomas}, which is most evident in optimally doped YBCO with $%
x=6.94$ and also for $x\sim 6.75.$ This long lived component is attributed
to quasiparticle relaxation in localized states and was discussed in detail
by Stevens et al\cite{Stevens}. In this paper we will only discuss the fast
component of the relaxation. The amplitude of this component was measured 
{\em well below }$T_{c}$ as a function of doping $x$ in YBCO over a wide
range of doping. The results shown in Figure 2 indicate that the integrated
amplitude - which is proportional to the total number of particles excited, $%
n=$ $\int (dn/dt)dt\,$ $\propto $ $\int_{-\infty }^{5\tau }\delta {\cal T}/%
{\cal T\,}dt\,$- increases almost linearly with doping, starting at the
insulator-to-metal transition at $x=6.4.$ (Near optimum doping, the
non-integrated amplitude of the signal actually decreases, but this is not
attributed to a drop in the DOS, but rather to the fact that the relaxation
time $\tau $ starts to increase dramatically for $x$ $\rightarrow 6.9.)$

An important feature of the data is that the amplitude is strongly
temperature dependent (Figure 3). At low temperatures ($T<T_{c}$) the
amplitude is nearly temperature independent. With increasing temperature the
amplitude starts to drop and eventually vanishes, such that at room
temperature the signal is usually not visible, or very small. The
temperature $T^{*}$ at which the amplitude $\delta {\cal T}/{\cal T}$ drops
to zero - indicating a change (drop) in the DOS\ at $E_{F}$ - increases with
decreasing carrier concentration in the underdoped phase in close agreement
with the ''pseudogap'' temperatures for YBCO determined from other
experiments\cite{NMR}. Importantly, near optimum doping the amplitude
vanishes near $T_{c}$, which has already been noted previously\cite{Stevens}%
. In underdoped samples on the other hand, the amplitude shows {\em no
change }at $T_{c},$ suggesting no changes in the DOS\ take place at $T_{c}$,
which as we shall discuss below is presented as evidence for Bose-Einstein
condensation.

As already mentioned briefly above, the quasiparticle lifetime $\tau _{s}$
in the superconducting state - determined by fitting a single exponential to
the time resolved trace - shows a divergence with increasing carrier
concentration $1-\delta ,$ in the region close to optimum doping as shown in
Figure 4. In contrast, almost no change in lifetime with $\delta $ is
observed above $T_{c}$, and is $\tau _{n}=0.5\pm 0.1$ ps at 100 K over the
whole range of doping. Importantly, the lifetimes below and above $T_{c}$ in
the underdoped state are typically the same below $x\sim 6.85.$ Approaching
optimum doping, this is no longer the case, and a clear divergence of $\tau $
is observed at $T_{c},$ behavior which is well known in BCS\ superconductors%
\cite{Rothwarf and Taylor} and is thought to be due to a quasiparticle
relaxation bottleneck as $T\rightarrow T_{c}$ from below (not a
Hebel-Slichter peak).

\section{Discussion}

The systematic investigation of the amplitude and relaxation time $\tau $ of
the photoinduced optical transmission through thin film samples shows rather
well the cross-over in behavior which occurs near optimum doping. In the
underdoped state, neither the amplitude (which is related to the DOS at $%
E_{F}$) nor the lifetime $\tau $ show any effect at $T_{c}$. Instead, both
show a gradual change with temperature, the amplitude vanishing at $T^{*}$,
while the lifetime showing only very gradual change up to $T^{*}$. The
simplest way to explain the behavior in the underdoped state is to assume
that no change in amplitude or relaxation time is observed at $T_{c}$
because no change in DOS (i.e. gap) occurs at $T_{c}$. Instead - in a
typical Bose-Einstein scenario - the pairs which are formed at $T^{*}$ only
acquire phase coherence at $T_{c}$.

The onset of a macroscopically coherent condensed state occurs when the
wavefunctions of adjacent pairs overlap sufficiently for phase coherence to
be established between them. We can estimate the temperature at which the
phase coherence is established by considering when the DeBroglie wavelength $%
\lambda $ becomes comparable to the superconducting coherence length $\xi .$
Thus $k_{B}T_{c}\simeq \hbar ^{2}/(m^{*}\xi ^{2}),$ which, using measured
values of $\xi =18$ A for YBaCuO\cite{Poole} and $m^{*}=3m_{e},$ gives $%
T_{c}=91$ K, which is close to the observed $T_{c}$-s in this material. The
density of carriers at the point of condensation is given by Einstein\cite
{Einstein} as $(N/V)^{2/3}=2.612^{-2/3}m^{*}k_{B}T_{c}/2\pi \hbar ^{2}$
which, using the same value of $m^{*}$ and $T_{c}$ as above, gives $%
n_{c}=N/V\sim 10^{21}cm^{-3}$ which is close to the value estimated from
Hall data\cite{Poole}.

The data in the near-optimally doped material with $x>6.85$ are distinctly
different to the undedoped state. Both the amplitude and the relaxation time
show a dramatic change at $T_{c},$ the latter exhibiting diverging behavior
as $T$ approaches $T_{c}$ similar to BCS superconductors like Al\cite
{Aluminium divergence}. The amplitude on the other hand shows a rapid drop
to zero very close to $T_{c}$, also with a BCS-like $T$-dependence,
suggesting that a gap opens simultaneously with pair formation (and phase
coherence) as in the BCS\ case.

To conclude, time-resolved optical spectroscopy is demonstrated to be a
powerful tool for the investigation of quasiparticle dynamics in HTS. The
occurrence of BE\ condensation in underdoped YBCO - as suggested by the
absence of any observable change in the quasiparticle relaxation time $\tau $
or density of states $N(0)$ at $T_{c}$ in our experiments - is consistent
with simple theoretical expectations. Obeyance of the sum rule implies that
pairs must form above $T_{c}$ (the observed changes in the DOS\ from $\Delta 
{\cal T}/{\cal T}$ suggest that this occurs at $T^{*}$). In contrast, the
optimally doped and overdoped state appears to show a divergence in the
quasiparticle lifetime with doping and a BCS-like temperature dependence of
both the intensity and lifetime is also apparent.

\section{Figure Captions}

Figure 1. A schematic diagram of the pump and probe processes in resonance
with the 1.5 eV CT\ transition in YBaCuO. The carriers relax rapidly to a
quasi-steady state between the pump and the probe pulses.

Figure 2. The amplitude of the induced transmission (squares)\ increases
with doping, following $T_{c}$. The data for $T_{c}$ are from Conder et al%
\cite{Conder Tc}.

Figure 3. The temperature dependence of the amplitude of the induced
transmission for a number of different samples. The $T_{c}$-s are shown in
the plot. The temperature at which $\Delta T/T$ drops to zero is close to
the pseudogap temperature $T^{*}$.

Figure 4. The relaxation time measured as a function of doping in the normal
state at 100 K (top panel) and at 20 K (lower panel). A clear divergence of $%
\tau _{s}^{20K}$ is observed near optimum doping.

\section{Acknowledgments}

We wish to acknowledge V.V.Kabanov for useful discussions and T.Mertelj for
his comments regarding the present manuscript.

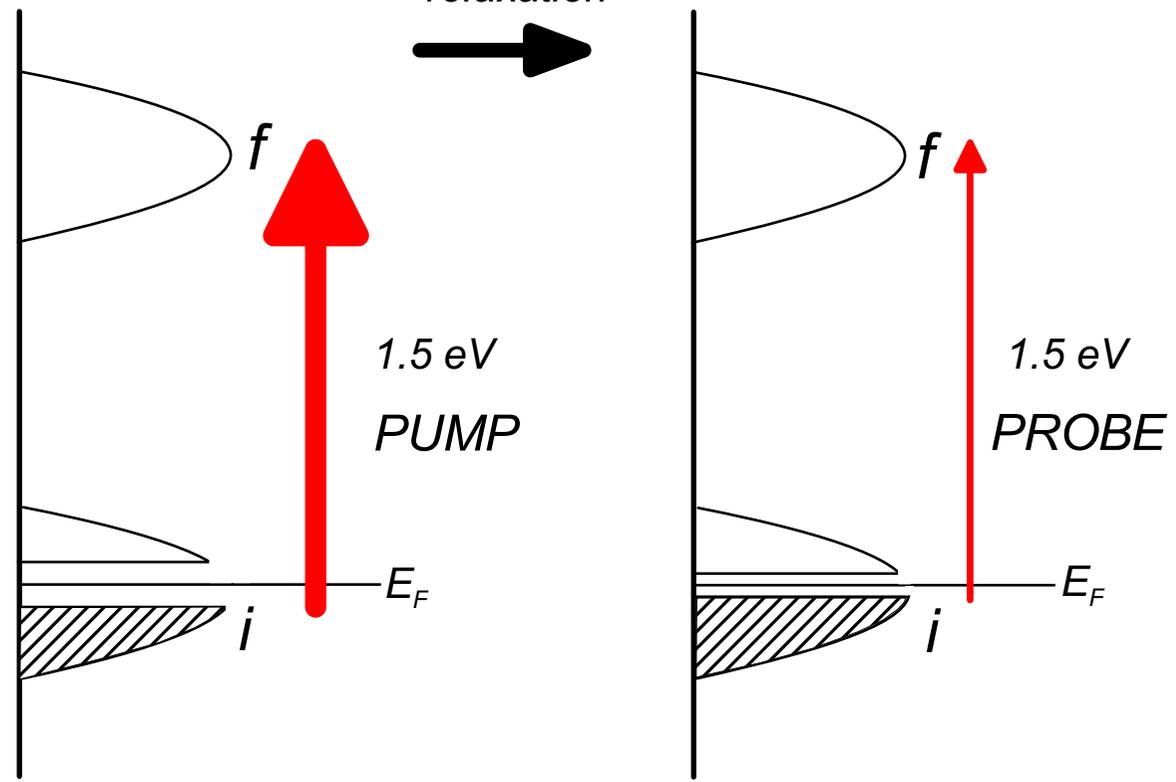
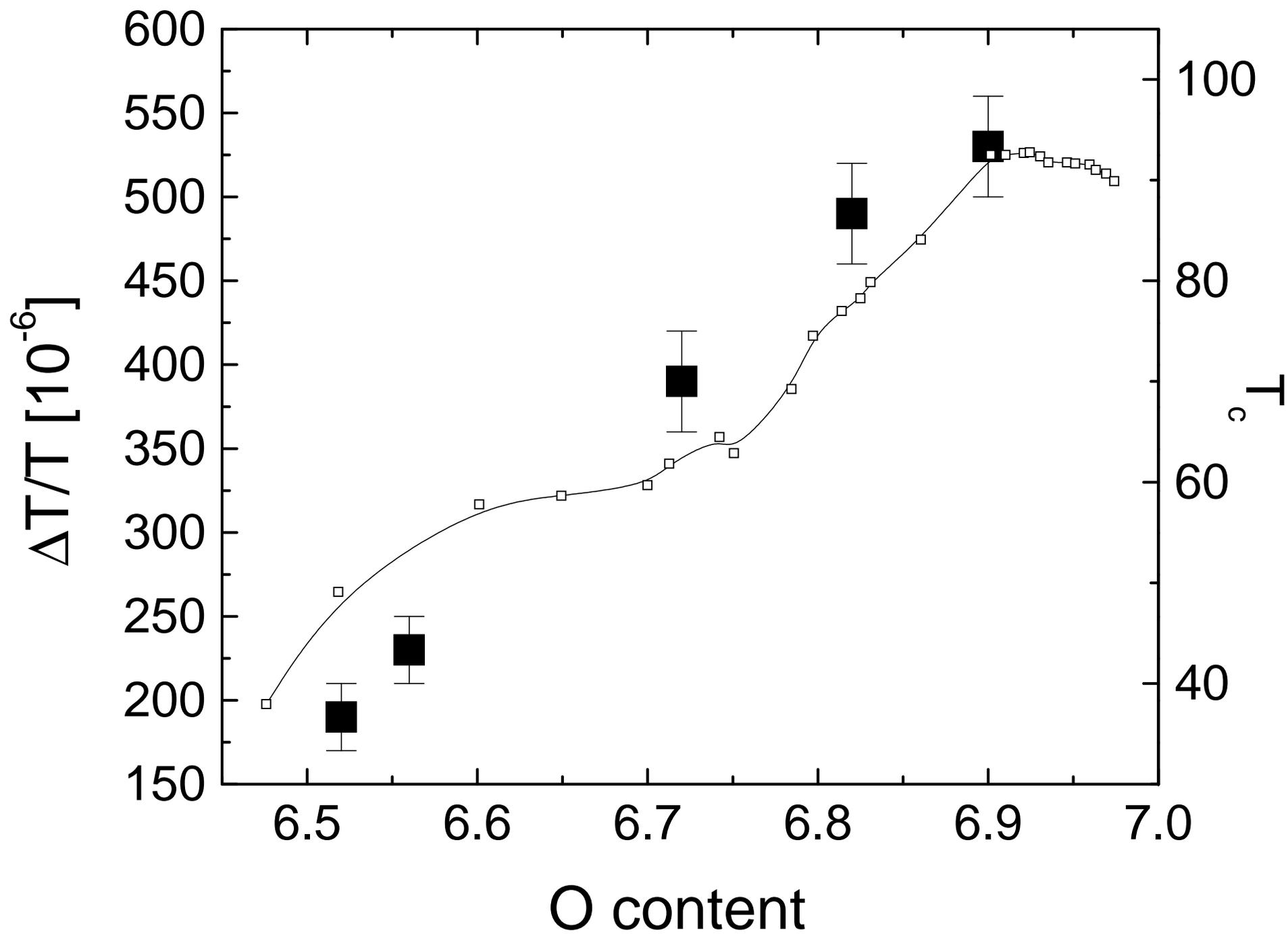
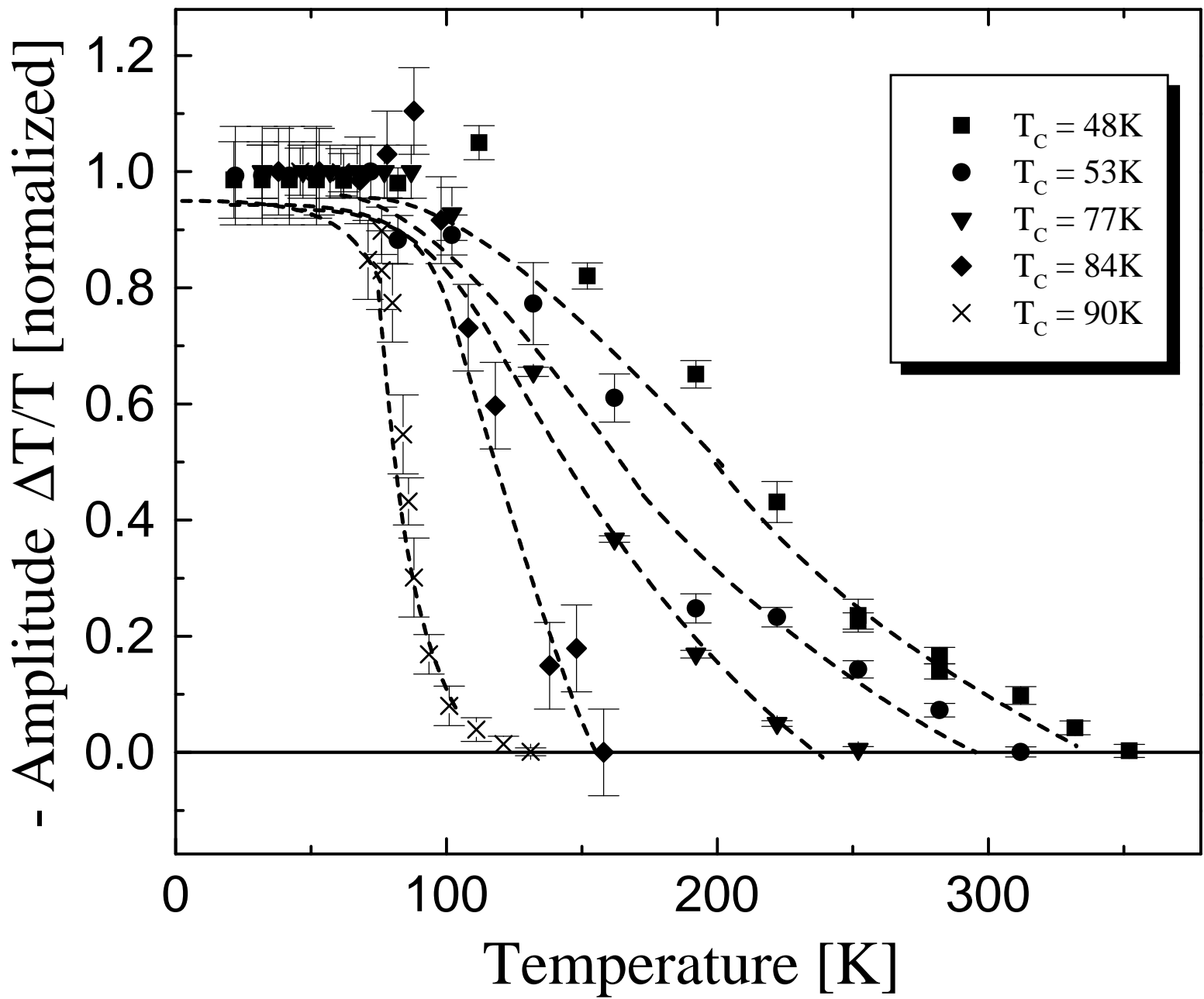
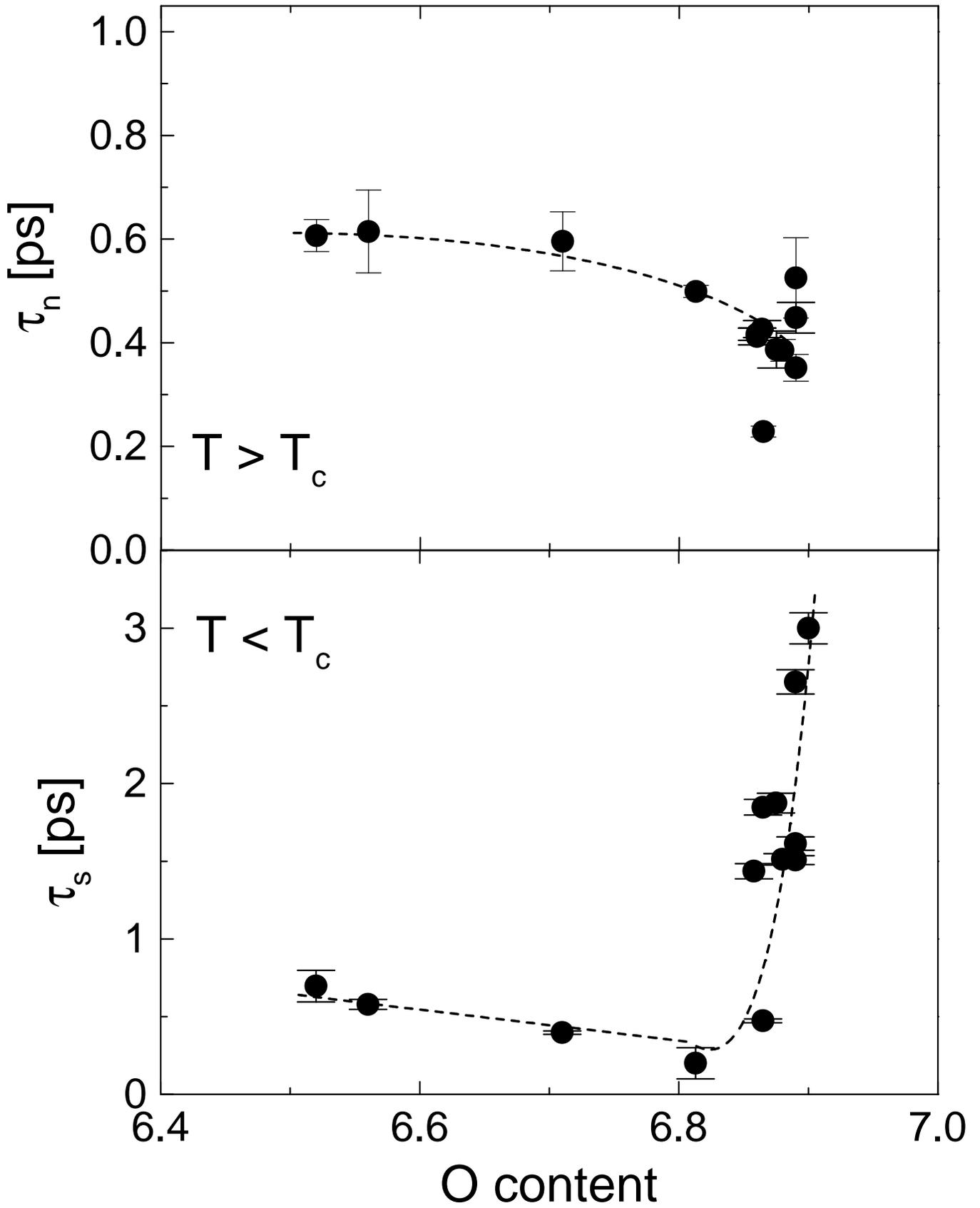
\end{document}